\documentstyle[graphicx,twocolumn,aps]{revtex}

 \newcommand{\etal}      {{\it et al}.}
 \newcommand{\cecoin}    {CeCoIn$_5$}
 \newcommand{\Ktperp}    {$^{115}K(2)_{\perp}$}
 \newcommand{\Ktpar}     {$^{115}K(2)_{||}$}
 
 \newcommand{\Koc}       {$^{115}K(1)_{c}$}
 \newcommand{\Koab}      {$^{115}K(1)_{ab}$}
 \newcommand{\KCoc}      {$^{59}K_{c}$}
 \newcommand{\KCoab}     {$^{59}K_{ab}$}
 \newcommand{\In}        {$^{115}$In}
 \newcommand{\Co}        {$^{59}$Co}
 \newcommand{\nuQ}       {$\nu_{Q}$}
 \newcommand{\tc}        {$T_{c}$}

 \title{ Anomalous NMR Magnetic Shifts in \cecoin }
 \author{N. J. Curro,$^1$ B. Simovic,$^1$ P. C. Hammel,$^1$ P. G. Pagliuso,$^1$
  G. B. Martins,$^2$ J. L. Sarrao,$^1$ J. D. Thompson$^1$}
 \address{$^1$Condensed Matter and Thermal Physics, Los Alamos National Laboratory, Los Alamos, NM 87545, USA\\
 $^2$National High Magnetic Field Lab and Florida State University, Tallahassee, FL 32306}

\address{
 \parbox{14cm}{\bigskip\rm\small
 We report \In\ and \Co\ Nuclear Magnetic Resonance (NMR)  measurements
 in the heavy fermion superconductor \cecoin\ above and below \tc.  The
 hyperfine couplings of the \In\ and \Co\ are anisotropic and exhibit dramatic changes below
 50K  due to changes in the crystal field level populations of
 the Ce ions. Below \tc\ the spin susceptibility is suppressed, indicating singlet pairing.
  \\ PACS Numbers: 74.70.Tx, 76.60.Cq }}

\begin{document}
 \maketitle

  \thispagestyle{myheadings}
  \markright{
  {\em LA-UR-01-1807}
  }

\narrowtext

In heavy fermion systems the interplay of magnetism and
superconductivity gives rise to a diverse range of ground states
including an unconventional form of superconductivity. The
recently discovered family of heavy fermion compounds CeMIn$_5$,
where M = Co, Rh or Ir exemplifies these effects. Whereas the Rh
compound undergoes a transition from antiferromagnetic to
superconducting under pressure \cite{helmut}, the Ir
\cite{petrovicIr} and Co \cite{petrovicCo} compounds superconduct
at ambient pressure, with the Co system exhibiting the highest
known transition temperature (2.3K) for any heavy fermion system.
Evidence from heat capacity, thermal transport and $\mu$SR
indicate that the pairing symmetry in the superconducting state is
unconventional and that there are line nodes in the
superconducting gap.\cite{roman,heffner}

The bulk magnetic susceptibility, $\chi$, of tetragonal CeMIn$_5$
displays systematic trends consistent with the diversity of
observed ground states. In all three cases $\chi$ is anisotropic,
and is largest for field applied along the $c$ direction. In the
$ab$ plane, $\chi_{ab}$ is essentially the same for all three
materials. However, $\chi_c$ exhibits a maximum at $\sim10$ K for
CeRhIn$_5$ ($ T_N = 3.8$ K), whereas for the superconductors
CeIrIn$_5$ and \cecoin\ $\chi_c$ diverges at low temperatures
until \tc\ is reached.  For both of these materials $\chi_c$ also
exhibits a plateau-like feature around 50 K, which is less
pronounced for the Ir system.  The origin of this feature and the
relationship between $\chi_c$ and \tc\ have been sources of
debate, however both the plateau and the divergence are intrinsic
and independent of field. \cite{petrovicCo}

Here we report a detailed study of site-specific magnetic shifts
in CeCoIn$_5$ using nuclear magnetic resonance (NMR).
Measurements in the normal state provide a microscopic measure of
the local susceptibility and we find anomalous temperature
dependencies. This behavior is likely due to the thermal
depopulation of a crystal field (CEF) excitation of the Ce ions.
We find remarkably strong departures from the expected
proportionality between bulk susceptibility and the NMR Knight
shift. We will argue that this effect is indicative of a high
degree Ce moment localization, a feature that may play a role in
the mechanism for superconductivity in this material.  In the
superconducting state the temperature dependencies of the shifts
reveal a suppression of the spin susceptibility consistent with
spin-singlet pairing.

Crystals of \cecoin\ were grown from an In flux as described in
\cite{petrovicCo}.  The tetragonal crystal structure of \cecoin\
consists of alternating layers of CeIn$_3$ and CoIn$_2$ and so
has two inequivalent In sites per unit cell. The In(1) site has
axial symmetry and is analogous to the single In site in cubic
CeIn$_3$. There are four low symmetry In(2) sites per unit cell,
two on each of the lateral faces of the unit cell, located a
distance 0.306$c$ above and below the Ce-In layer.
\cite{structureref} The zero field \In\ ($I=9/2$) Nuclear
Quadrupolar Resonance (NQR) spectrum reveals an axially symmetric
site with $^{115}$\nuQ(1)$=8.173\pm0.005$ MHz, and $\eta(1)=0.0$
at 4 K, whereas the electric field gradient (EFG) at the In(2)
site is characterized by $^{115}$\nuQ(2)$=15.489\pm0.001$ MHz, and
$\eta(2)=0.386\pm0.001$, where \nuQ\ and $\eta$  are defined as in
\cite{CeRhIn5paper,CPSbook}. The NMR spectrum of the \Co\
($I=7/2$) indicates a site with axial symmetry and $^{59}\nu_Q =
234\pm$1 kHz at 4 K.  Both the In and the Co EFG's are
essentially temperature independent, varying less than 0.5\%
between 4 K and 100 K, indicating that significant structural
changes are absent in this temperature range.

The magnetic shift measurements were made on a large single crystal of
\cecoin, which was mounted with the $c$ axis either parallel or
perpendicular to the external field, for fields between 3 and 5 T.
Field-swept spectra were obtained by measuring the spin echo intensity
as a function of applied field at fixed frequency. The shifts were
determined by measuring several of the \In\ transition fields $H_{\rm
exp}$ for each site at several different fixed frequencies. The nuclear
  spin Hamiltonian $ {\mathcal H} =
(h\nu_Q/6)[3I_z^2-I^2+\eta(I_x^2-I_y^2)] +\gamma\hbar{\mathbf
I}\cdot({\mathbf 1}+\mathbf{K})\cdot{\mathbf H}_0$, where ${\mathbf
K}=(K_a,K_b,K_c)$ is the magnetic shift tensor, was diagonalized and
the resonance fields $H_{\rm res}$ for each transition and each In site
were then calculated. The spectra were then fit by minimizing
$\chi^2=\sum_i(H_{\rm res} - H_{\rm exp})^2$ as a function of
$(\theta,\phi,K_a,K_b,K_c)$, where $\theta$ and $\phi$ are the polar
angles relating ${\mathbf H}_0$ to the crystal axes $(a,b,c)$.  Note
that such a procedure is necessary because the strong quadrupolar
interaction gives rise to a significant angular dependence of $H_{\rm
res}$ so that even a misalignment of 1-2$^{\circ}$ can cause a
significant error ($\sim 30\%$) in ${\mathbf K}$. The Co shift and EFG
were determined by measuring the positions of the central and satellite
transitions at fixed field.

Given three nuclei and two possible field orientations for each
there are seven distinct magnetic shifts. Note that for In(1) and
Co the magnetic shift is isotropic in the $ab$ plane, whereas for
In(2) the shift differs for ${\mathbf H}_0$ parallel or
perpendicular to the unit cell face. The temperature dependencies
of $K$ for both In sites as well as the Co are shown in Fig. (1),
together with $\chi$ for both directions. $K$ is a measure of the
local electronic spin density at the nuclear site.  In general,
the shift is given by: $K(T)=K_0+\sum_iA_i\chi_i(T)$, where $K_0$
is an orbital shift, independent of the local spin density at the
nuclear site and the temperature, and $A_i$ is the hyperfine
coupling to $\chi_i$, the $i^{th}$ component of the
susceptibility $\chi=\sum_i\chi_i$. Both $K_0$ and $A_i$ can be
anisotropic. All of the magnetic shifts except \Ktperp\ are
proportional to $\chi$ for $T\gtrsim40$K for ${\mathbf H}_0 ||
ab$  and $T\gtrsim60$K for ${\mathbf H}_0 || c$. Below these
temperatures \Ktpar,\KCoab, \KCoc\ and \Koc\ show dramatic
departures from $\chi$. Furthermore, \Ktperp\ is not proportional
to $\chi$ in any temperature regime, and exhibits a dramatic
downturn below 40 K   [note that the axis for \Ktperp\ is
reversed in Fig. (1)]. Figure (2) shows $K$ versus $\chi$ for both
field directions.
\begin{figure}
  \centering
  \includegraphics[width=\linewidth]{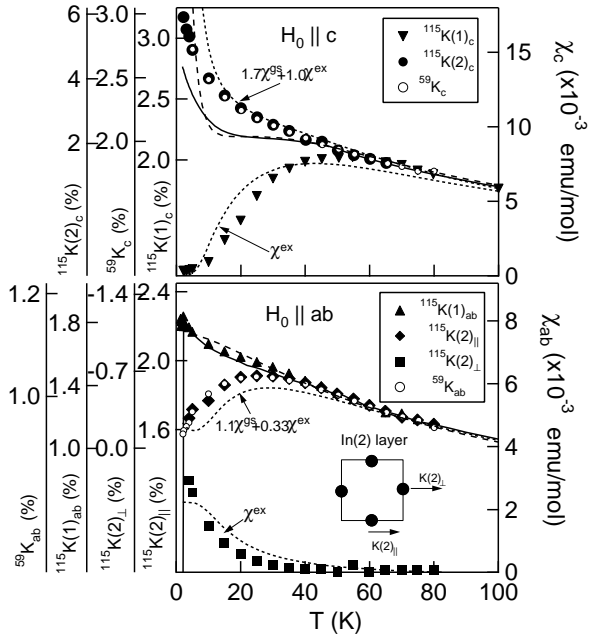}
  \caption{The magnetic shift versus temperature.
  The solid lines show $\chi_c$ and $\chi_{ab}$. The dashed line is a
  fit to $\chi$ as described in the text; the dotted lines,
  which have been offset vertically to coincide with the shift scales, are
  calculations as described in the text.  [Note the reversed axis for \Ktperp.]}
  \label{fig:1}
\end{figure}
\noindent Note that $K\propto \chi$ for high temperatures ($T>40$
K), and the intercept and slope give $K_0$ and $A$, whose values
are listed in Table (1), where $A_{HT}$ is determined for high
temperatures, and $A_{LT}$ for low temperatures.  The Co shifts
track those of the In(2) for both directions, where $A_c({\rm
Co})/A_c({\rm In(2)})=0.26$ and $A_{ab}({\rm Co})/A_{\perp}({\rm
In(2)})=0.33$. Therefore, it seems likely that the Co is not
directly coupled to the Ce, but couples to the Ce only via the
In(2).

Anomalous departures from $K\propto\chi$ have been known to exist
in Ce compounds for several years, although the reason for the
departure is still under debate. \cite{maclaughlin,cox,ohama} It
is generally considered that the Ce 4f electron does not have a
significant direct overlap with the orbitals of neighboring
nuclei.  Rather, it is the 6s and 5d orbitals of the Ce that are
hybridized, and the Ce 4f moment can create a hyperfine field at
a neighboring atom by polarizing the conduction electrons at the
Ce site, which is then transferred to the neighbor via an RKKY
interaction. The conduction electrons at the neighbor then create
a hyperfine field at the nucleus via a contact interaction.  Two
different mechanisms have been proposed to explain the anomalous
shift behavior in other heavy fermion systems. In
  CeSn$_3$ $K$(Sn) and $\chi$ differ below $\sim 150$ K,
and this effect has been ascribed to modifications of the effective
hyperfine coupling at the Sn (via the RKKY interaction) by the onset of
Kondo compensation below a temperature $T_K$. \cite{cox,malik} In
CeCu$_2$Si$_2$ Ohama and coworkers observed that the Cu and Si magnetic
shifts also exhibit departures from $K\propto\chi$ below $\sim 100$ K,
and they attribute this behavior to the depopulation of an excited CEF
level of the Ce ions ($J=5/2$) and not Kondo coherence.\cite{ohama} In
this case, the overlap between the Ce 4f orbitals and the conduction
electrons differs depending on the CEF level populations, resulting in
temperature dependent hyperfine couplings to the Cu and Si.
\begin{figure}
  \centering
  \includegraphics[width=\linewidth]{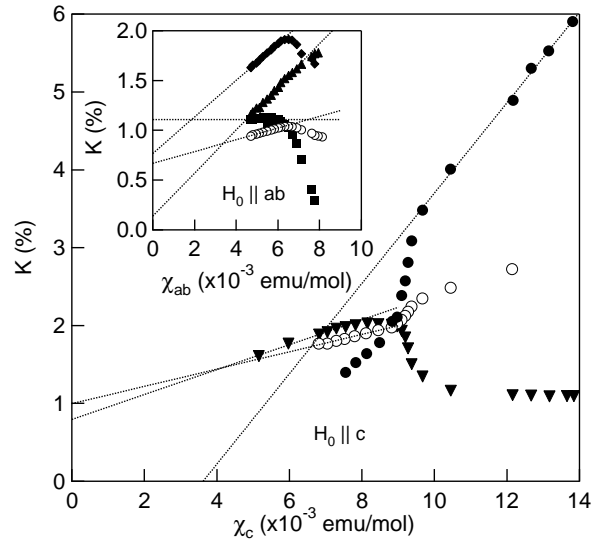}
  \caption{$K$ versus $\chi$ in both directions.
  The dotted lines are linear fits to the high temperature data,
  and the symbols are the same as in Fig. (1).}
  \label{fig:2}
 \end{figure}
\noindent In fact, the measured shifts in \cecoin\ show behaviors
similar to those observed in CeCu$_2$Si$_2$. Namely, the $K$
versus $\chi$ plots exhibit positive slope at high temperatures;
however, at low temperatures $K\propto\chi$ is recovered, but
with a negative slope.  Ohama \etal\ attribute the negative
hyperfine coupling to an orbital overlap between the ligand s
orbital and the Ce 4f orbital. They distinguish this direct
transferred hyperfine mechanism from that in which the 4f moment
polarizes the conduction band at the Ce site. According to Ohama
\etal, the direct contribution can become negative when only the
lowest CEF doublet is occupied. Heat capacity data in \cecoin\
suggest the presence of an excited CEF doublet at $\sim$ 50K
above the ground state doublet, \cite{petrovicCo} so it would be
reasonable to ascribe the anomalous shift behavior in \cecoin\ to
depopulation of an excited CEF doublet. The strong site
dependence of the shift anisotropy in \cecoin\ also suggests a
direct coupling between the In or Co and an anisotropic,
localized Ce 4f orbital, as in CeCu$_2$Si$_2$.  Recent deHaas van
Alphen and photoemission data in \cecoin\ are consistent with LDA
calculations, which assume the Ce 4f electron is itinerant.
\cite{dHvAandPES} In contrast, the strong directionality of the
hyperfine couplings in these materials implies an electronic
structure that is more tight binding rather than free electron
like. In both systems the temperature scale for Kondo
compensation is much lower than the CEF splitting, suggesting
that more localized 4f character, which could be a source for
spin fluctuations, is essential for the development of heavy
fermion superconductivity.

In order to investigate the possible role of CEF effects we have
fit $\chi$ to extract the CEF parameters.  The dashed lines in
Fig.~(1) show a fit to the expression
$\chi^{-1}=\chi_{\text{CEF}}^{-1}+\lambda$, where
$\chi_{\text{CEF}}$ is the CEF susceptibility for the Ce ion, and
$\lambda$ is a molecular field term.  The Ce ion in \cecoin\
experiences a crystal field with tetragonal symmetry, so
 ${\mathcal H}_{\text{CEF}} =
b_2^0O_2^0+b_4^0O_4^0+b_4^4O_4^4$, where the $O_n^m$ are the
Steven's operators. \cite{stevens}  In this field the $J=5/2$
manifold is split into three doublets
$(\Gamma_6,\Gamma_7^{(1)},\Gamma_7^{(2)})$, where the
wavefunctions are given by: $\left|\pm\frac{1}{2}\right>$,
$\mp\sin{\alpha}\left|\pm\frac{3}{2}\right>\mp\cos{\alpha}\left|\mp\frac{5}{2}\right>$,
$\mp\cos{\alpha}\left|\pm\frac{3}{2}\right>\pm\sin{\alpha}\left|\mp\frac{5}{2}\right>$,
and $\chi_{\text{CEF}}=(\partial^2(\log{Z})/\partial H^2)_{H=0}$.
Here $Z$ is the partition function for the Hamiltonian ${\mathcal
H}_{\rm Ce}={\mathcal H}_{\text{CEF}}+g_J\mu_B{\mathbf
H}\cdot{\mathbf J}$, where $g_J=6/7$, $\mu_B$ is the Bohr
magneton, and ${\mathbf J}$ is the spin operator for $J=5/2$. We
find the best fit for the $\Gamma_6$ ground state
($J_z=\pm\frac{1}{2}$), with excited states at 34K and 102K above
the ground state, $\alpha=1.47$, and an anisotropic molecular
field: $\lambda_c=18.8$ mol/emu and $\lambda_{ab}=-113.2$
mol/emu. \cite{CEFnote} The anisotropy of $\lambda$ reflects Ce-Ce
couplings which differ for neighbors in and out of the plane. The
fit reproduces the plateau feature, and suggests that the
anomalous behavior of the magnetic shifts below 50K may also be
explained by changes in the hyperfine couplings as the excited
CEF states are depopulated. Note, for example, that in Fig. (2)
\Koc\ appears to be independent of $\chi_c$ at low temperatures.
This behavior suggests that \Koc\ couples only to the excited CEF
states. If we decompose $\chi_{\text{CEF}} =
\chi_{\text{CEF}}^{\text{gs}} + \chi_{\text{CEF}}^{\text{ex}}$
into contributions from the ground state doublet and from the
excited doublets, one might then expect
$K=K_0+A_{\text{gs}}\chi^{\text{gs}}+A_{\text{ex}}\chi^{\text{ex}}$,
where $(\chi^{i})^{-1}=(\chi_{\text{CEF}}^{i})^{-1}+\lambda$. We
determine $\chi_{\text{CEF}}^{\text{gs}}$
($\chi_{\text{CEF}}^{\text{ex}}$) by suppressing the field
dependence of the excited (ground) state energy levels in the
expression for $Z$. By adjusting $K_0$, $A_{\text{gs}}$ and
$A_{\text{ex}}$ appropriately, we can qualitatively explain the
temperature dependence of all the shifts in Fig. (1) (dotted
lines). Note that $\chi_{\text{CEF}}^{\text{ex}}<0$ in the $ab$
plane, so for this component the absolute value of $\lambda_{ab}$
was used.

The anomalous behavior of the shifts might also be explained by
two components of $\chi$ with a different origin than crystal
field states.  However, there is only one Ce site in the unit
cell, and susceptibility and heat capacity data indicate that the
observed properties can be entirely attributed to the Ce (i.e., Co
is nonmagnetic in \cecoin).  Therefore it seems likely that the
two components can only be attributed to different CEF states on
the Ce ions. It is interesting to note that measurements of the
In(1) shift in the isostructural compound CeRhIn$_5$ reveal a
positive hyperfine coupling for 4 K $< T <$ 50 K, with no signs
of the dramatic departure from $K\propto\chi$ seen in
\cecoin.\cite{Rh115shift}  Clearly, if the hyperfine anomaly is
the only mechanism at work in \cecoin\ then the CEF parameters in
\cecoin\ must differ significantly from those in CeRhIn$_5$. In
fact, recent work by Takeuchi and coworkers suggests that the
ground state CEF level in CeRhIn$_5$ is $\Gamma_7$ rather than
$\Gamma_6$. \cite{onuki}

\begin{figure}
  \centering
  \includegraphics[width=\linewidth]{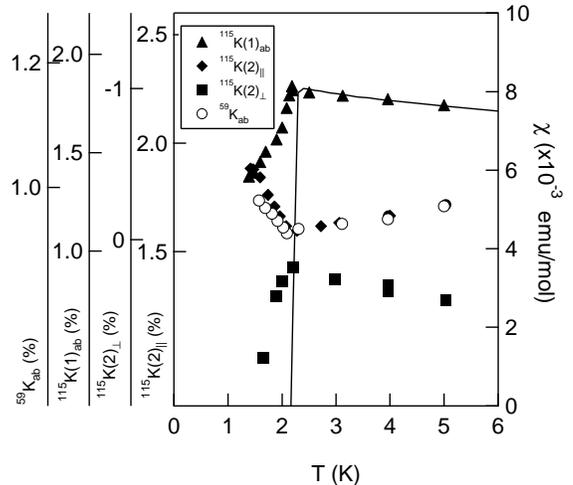}
  \caption{The magnetic shift for ${\mathbf H}_0 || ab$ in the superconducting state.
   The solid line is the bulk susceptibility,
   which becomes fully diamagnetic below 2.3 K.}
  \label{fig:4}
\end{figure}

Below \tc\ $\chi$ is dominated by the diamagnetic response,
masking the intrinsic behavior of the spin susceptibility.  $K$,
however, couples only to the spin susceptibility and provides a
direct measure of $\chi^{\rm spin}$ in the superconducting state.
The temperature dependencies of the shifts for both In sites as
well as the Co in \cecoin\ are shown in Fig. (3) for ${\mathbf
H}_0 || ab$ down to 1.4K . Because of the thin platelet
morphology of \cecoin, demagnetization fields in the
superconducting state can be significant for ${\mathbf H}_0 ||
c$, precluding an accurate determination of the magnetic shift
since the local field at the nucleus is poorly determined.  We
estimated that for our sample, the demagnetization factor for $
{\mathbf H} || c$ is $N_c/4\pi\approx0.79$. Therefore, although
we observe a decrease in the resonance frequencies for this
direction one cannot resolve whether the decrease is due to a
change in $K$ or to a change in $H_0$ internally. However, for
${\mathbf H}_0 || ab$ the demagnetization factor is much smaller,
so the internal field below \tc\ is known to a greater degree of
accuracy. Therefore, we only present data on the shifts for the
field in the plane.

\begin{figure}
  \centering
  \includegraphics[width=\linewidth]{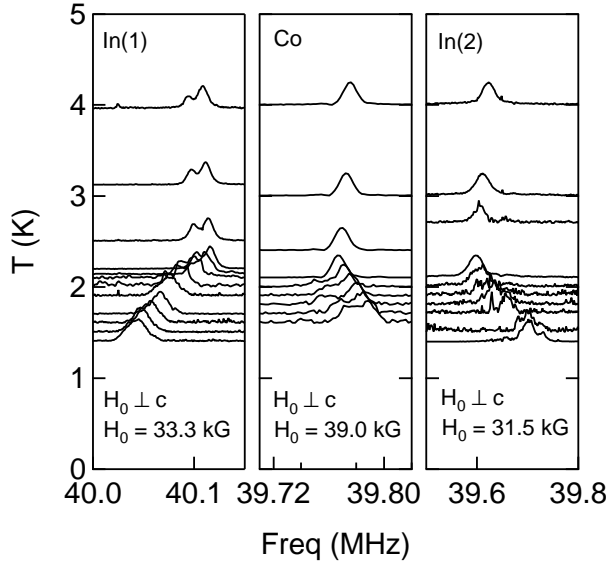}
  \caption{Spectra of the In(1), Co and In(2) at various temperatures
  through \tc. The In(2) spectrum is for ${\mathbf H}_0$ normal to the
  unit cell face. The two peak structure of the In(1) spectrum indicates
  the presence of two slightly differently oriented crystals in the sample.}
  \label{fig:5}
\end{figure}

The decrease in \Koab\ seen in Fig. (3) implies a decrease in
$\chi^{\rm spin}$. However, \KCoab, \Ktperp\ and \Ktpar\ {\it
increase} below \tc\ [note the reversed axis for \Ktperp\ in Fig.
(3)]. Spectra of the In(1), In(2) and Co at different temperatures
are shown Fig. (4), clearly exhibiting the behaviors seen in Fig.
(3).  An increase of the absolute value of \Ktperp, \Ktpar, and
\KCoab\ below \tc\ can be understood by recognizing that the
hyperfine coupling is negative below 50 K [see Fig. (3)], so an
increase in $K$ implies a decrease in $\chi^{\rm spin}$. Thus,
all of the shifts for ${\mathbf H\perp c}$ are consistent with a
decrease in $\chi^{\rm spin}$, implying spin-singlet pairing of
the Cooper pairs in the superconducting state.  Given the recent
heat capacity and thermal conductivity measurements revealing
higher orbital symmetry,\cite{roman,Izawa} we can conclude that
the order parameter in \cecoin\ has d-wave symmetry. During the
course of this work, we became aware of similar work by the group
of Kohara \cite{koharashift} who report magnetic shift results
below \tc. Although our conclusions about singlet pairing are the
same, the temperature dependencies of the shifts differ.

We thank S. Dunsiger for assistance with the measurements, as
well as D. MacLaughlin, R. Heffner, and J. Lawrence for valuable
discussions. This work was performed under the auspices of the US
Department of Energy.

\begin{table}
\caption{The hyperfine couplings and orbital shifts of the In(1),
In(2) and Co. }
\begin{tabular}{cccc}
  shift$_{\alpha}$ & $K_0(\%)$ & $A_{HT}$ (kOe/$\mu_{\rm B}$)  & $A_{LT}$ (kOe/$\mu_{\rm B}$) \\
  \hline
  In(1)$_c$ & 0.79(5) & 8.94(34)& -0.4(1) \\
  In(1)$_{ab}$ & 0.13(4) &12.08(40) & 12.08(40) \\
  In(2)$_c$ &  -2.10(3) & 32.4(3) & 22.8(3) \\
  In(2)$_{||}$ & 0.76(2) & 10.26(17) & -12(1)\\
  In(2)$_{\perp}$ & 1.10(1) & 0 & -34.7(9)\\
  Co$_c$ & 1.00(1) & 8.4(5)  & 6.20(5) \\
  Co$_{ab}$ & 0.68(1) & 3.30(9) & -4.20(19) \\
\end{tabular}
\end{table}


\begin{thebibliography}{29}
\bibitem{CEFnote} In terms of the parameters in ${\mathcal
H}_{\text{CEF}}$ we have $b_2^0=$5.614 K, $b_4^0= -0.0035 $ K, and
$b_4^4 = 0.260 $ K.
\end{thebibliography}
\end{document}